# Agent Based Software Testing Framework (ABSTF) for Application Maintenance

K. Karnavel[1], V. Divya[2], Gnanakeerthika[3] and P. Karthika[4]

[1]Assistant Professor, Department of Computer Science & Engg,
Anand Institute of Higher Technology, Chennai

[2,3&4] UG Student, Department of Computer Science & Engg,
Anand Institute of Higher Technology, Chennai

**Abstract**

Software testing framework can be stated as the process of verifying and validating that a computer program/application works as expected and meets the requirements of the user. Usually testing can be done manually or using tools. Manual testing can be time consuming and tedious, also it requires heavy investment of human resources. Testing tools in fact have many advantages and are widely used nowadays, but also has several disadvantages. One particular problem is that human intervention is no longer needed. Testing tools are of high cost and so, it cannot be used for small companies. Hence in this paper we propose Agent based testing, which is a fast growing approach in testing field. The proposed system (ABSTF) has to reduce the application testing moment, easily find out bug and solve the bug by Regression Testing. Here by we are going to use a safe efficient regression selection technique algorithm to selectively retest the modified program. We also use Traceability relation of completeness checking for agent oriented system, which identifies missing elements in entire application and classifies defect in the testing moment. With the ability of agents to act autonomously, monitoring code changes and generating test cases for the changed version of code can be done dynamically.

**Keywords:** *Regression Testing, Selection Technique Algorithm, Agent based software testing*

## 1. Introduction

Software engineering is the application of a systematic, disciplined, quantifiable approach to the design, development, operation, and maintenance of software, and the study of these approaches. According to Ian Somerville "Software testing is more than just error detection. Testing software is operating the software under controlled conditions, to verify that it behaves as specified to detect errors, and to validate that what has been specified is what the user actually wanted".

The Agent based system serves as a tool for the maintainers to improve the efficiency of software maintenance. This system will allow the maintainer to maintain the software while learning the application [3]. The main goal is that, if any changes have been made to the source code, the system will provide complete information about those changes which will allow the maintainers to check if the program works as desired.

The traceability relation technique is used to trace the bug in requirement, code and design of the software application, since the application is under constant change [5]. Then the changed part of the code is selectively retested using regression selection technique algorithm.

This system reduces the amount of work that the maintainers have to perform in the maintenance of the software application. It will also help reduce the time lost when current software designers hand off the project to maintainers. From the perspective of the experienced or relatively new software maintainer, the use of intelligent agents is a time saving tool for the software maintenance process.

## 2. Regression testing

Regression testing is any type of software testing that seeks to uncover new software bugs, in existing functional and non-functional areas of a system after changes, such as enhancements, have been made to them.[1]

The intent of regression testing is to ensure that a change such as those mentioned above, has not introduced new faults. One of the main reasons for regression testing is to determine whether a change in one part of the software affects other parts of the software.

## 2.1 Traceability relation

Traceability relation is a type of dependency relation between elements. An important activity in the development of software systems is the creation of traceability relation between the various models. A requirements traceability matrix may be used to check to see if the current project requirements are being met.

Traceability relations can assist with several activities of the software development process such as evolution of software systems, compliance verification of code, reuse of parts of the system, requirements validation and identification of common aspects of the system [7]. A traceability matrix is a table that correlates requirements or design documents to test documents. It is used to change tests when related source documents are changed, to select test cases for execution when planning for regression tests by considering requirement coverage.

## 2.2 Selection Technique Algorithm

Selective retest technique attempts to reduce the time required to retest a modified program by selectively reusing tests and selectively retesting the modified program [2]. The selection technique chooses, from an existing set of test cases, tests that are necessary to validate the modified program. We prove that under certain conditions, the set of test that our technique selects includes every test from the original test suite that can expose faults in the modified program.
This technique addresses two problems: 1. Problem of selecting test from an existing test suite and 2. The problem of determining where an additional test may be required. This algorithm is more general than most other techniques. It also handles all language constructs and all types of program modifications for procedural languages.

## 3. Problem Definition

### 3.1. Existing system

The existing system tries to increase the efficiency of regression testing. The regression testing paradigms have been recently devised using a number of automated testing tools, object orientation, component ware either to make the engineering process easier or to extend the complexity of applications that can feasibly be built[4].

In the existing system the software application testing is inefficient and very expensive. When user tests one software application the errors will not be traced easily. Human Testing effort will be high.

### 3.2. Proposed system

Proposed agent oriented software testing can be identified as a next generation model for engineering complex, distributed system[4]. The proposed system has to reduce the application testing moment, easily can find out bug and solve the bug by a safe efficient Regression selection technique [8].

A software agent is to provide the software maintainers with detailed information about any changes that have been made to the source code by any users [14] [2]. This information will allow the maintainers to check that the application is working as desired.

## 4. Design Procedure

### 4.1. Module Description

#### 4.1.1 User Input Module

In this module user imports the software file/package and testing process will be performed to find the missing elements and trace the bugs in the application.

#### 4.1.2. Traceability Agent Module

In this module a rule-based approach is used to support automatic generation of traceability relations and identification of missing element in developing software product or application and also finding the bugs.

### 4.1.3 Monitor Agent Module

Essentially the monitor agent acts as a trigger and sets the other agents in the system into motion when changes are found in the code. Otherwise, the agent will stay and continue to monitor the code base.

### 4.1.4. Impact Agent Module

In this module the agent framework (ABSTF) is to receive the information from the monitor agent and determine the impact of the changes made within the source code. Its specific goal is to search for the number of function calls to the changed functions in the source code.

### 4.1.5 Generating Agent Module

This module receives the source code changes information from impact agent and automate the test case generation for the changed code.

## 5. Agent Based Software Testing Framework (ABSTF)

The user imports the package which is then tested module by module using v-cycle. If no bugs are found test cases will be generated. Incase bugs are found, the location of the bug is found using traceability matrix, which is then cleared and retested using regression testing, then the test cases will be generated for the modified code and the output will be displayed (fig 1).

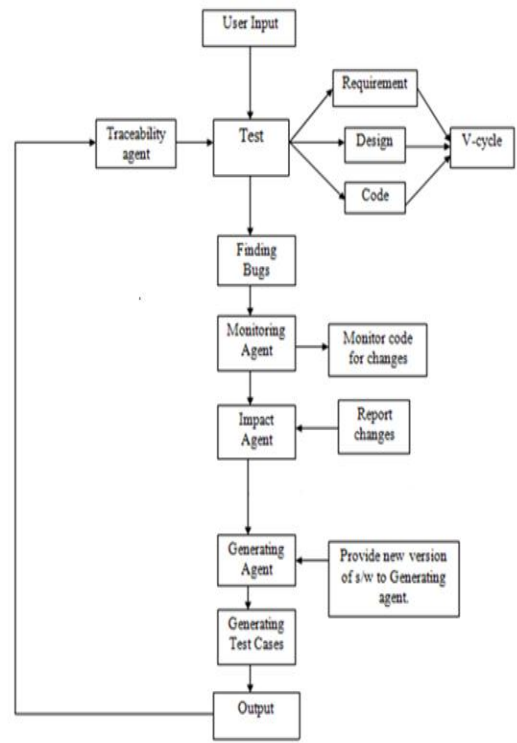

**Fig: 1**

## 6. Conclusion

Software Agent technology has drawn much attention as the preferred architectural technique for the design of many distributed software systems. Agent based systems are often featured with intelligence, autonomy, and reasoning [1]. In this paper we propose an agent based technique for automating the regression testing. To make a test case more adaptable to the change of application, we define an agent framework(ABSTF) based technique to generate test cases automatically. Here we use JADE which is a software Framework fully implemented in Java, used to test the application and agent based technique to generate test case which facilitates test case creation, execution and repair.

## 7. Future Enhancement

In future we are aiming to record the new errors in a database so that we may easily correct the errors without repeating the whole process. Apart from this we also plan to support all language based applications in future.